\documentclass[preprint,floats]{revtex4-1}

\usepackage{dcolumn} % Align table columns on decimal point
\usepackage{bm} % bold math
\usepackage{amsthm}
\usepackage{latexsym}
\usepackage{float}
\usepackage{amsfonts}
\usepackage{amsmath}
\usepackage{amssymb}
\usepackage{color,graphicx}
\usepackage{adjustbox}

\begin{document}

\title{Exciton Dynamics in Monolayer Transition Metal Dichalcogenides}

\author{Galan Moody}
\email{galan.moody@gmail.com}
\affiliation{National Institute of Standards $\&$ Technology, 325 Broadway, Boulder, CO 80305}
\author{John Schaibley}
\affiliation{Department of Physics, University of Washington, Seattle, Washington 98195}
\author{Xiaodong Xu}
\email{xuxd@uw.edu}
\affiliation{Department of Physics, University of Washington, Seattle, Washington 98195}
\affiliation{Department of Materials Science and Engineering, University of Washington, Seattle, Washington 98195}

\begin{abstract}
Since the discovery of semiconducting monolayer transition metal dichalcogenides, a variety of experimental and theoretical studies have been carried out seeking to understand the intrinsic exciton population decay and valley relaxation dynamics. Reports of the exciton decay time range from hundreds of femtoseconds to ten nanoseconds, while the valley depolarization time can exceed one nanosecond. At present, however, a consensus on the microscopic mechanisms governing exciton radiative and non-radiative recombination is lacking. The strong exciton oscillator strength resulting in up to $\sim20\%$ absorption for a single monolayer points to ultrafast radiative recombination. However, the low quantum yield and large variance in the reported lifetimes suggest that non-radiative Auger-type processes obscure the intrinsic exciton radiative lifetime. In either case, the electron-hole exchange interaction plays an important role in the exciton spin and valley dynamics. In this article, we review the experiments and theory that have led to these conclusions and comment on future experiments that could complement our current understanding.
\end{abstract}

\maketitle
\section{Introduction}
\begin{figure}[htbp!]
\centering
\includegraphics[width=0.4\linewidth]{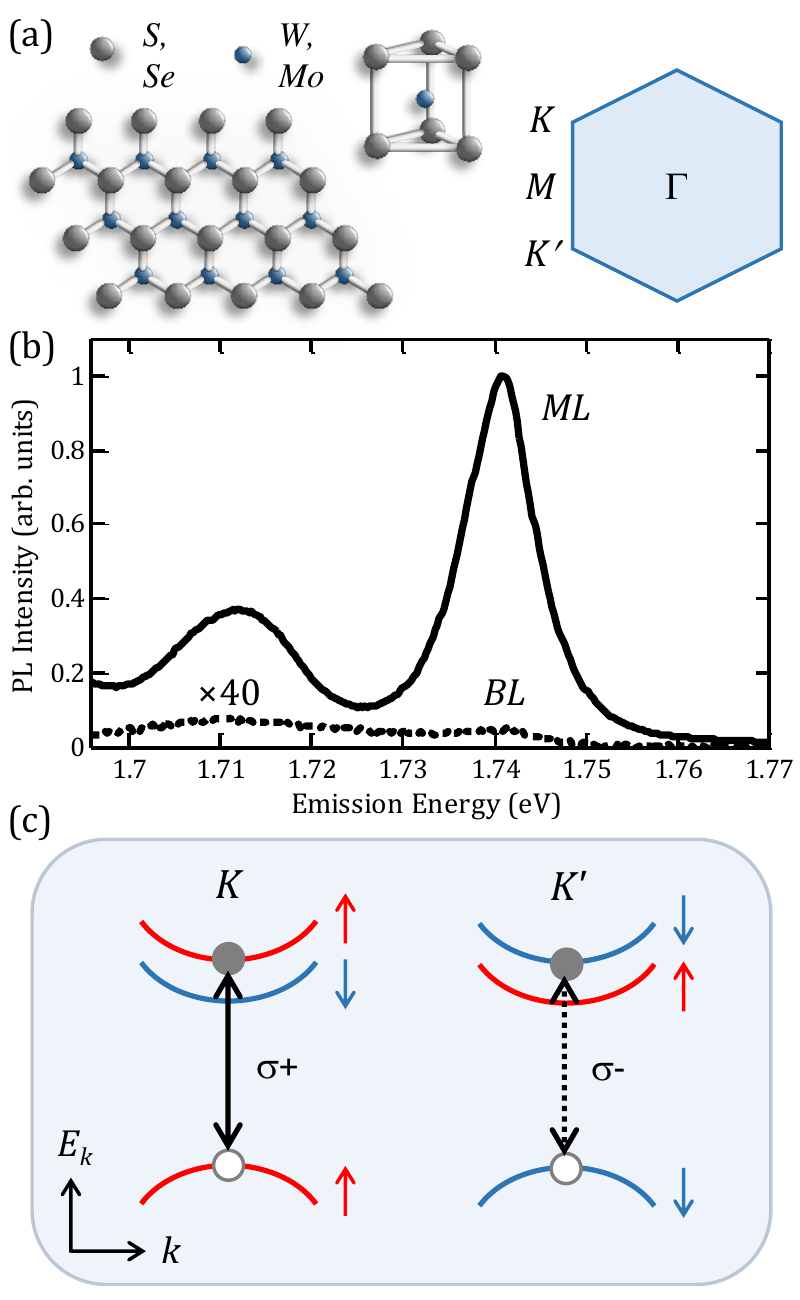}
\caption{(a) Schematic illustration of the two-dimensional hexagonal lattice formed from $\{Mo,W\}$ and $\{S,Se\}$ atoms. Optical band gaps form at the $K$ and $K'$ valleys on the edge of the Brillouin zone. (b) Photoluminescence spectrum for monolayer (ML) and bilayer (BL) WSe$_2$. (c) Sketch of the valley-dependent optical selection rules near the band edges. The electronic spin states are labeled by the arrows, shown for WX$_2$. In the $K$ ($K'$) valley, the optically active transition couples to $\sigma+$ ($\sigma-$) polarized light at normal incidence.}
\label{Fig1}
\end{figure}

Two-dimensional materials such as graphene, black phosphorous, and transition metal dichalcogenides (TMDs) exhibit fascinating physical properties stemming from their unique band structure and reduced dimensionality \cite{Novo2005}. In recent years, group-VIB TMDs ($MX_2$, where $M = Mo, W$ and $X = S, Se$) in particular have attracted significant interest as a novel testbed of exciton many-body physics of 2D systems \cite{Qiu2013}, while providing an excellent platform for ultrathin optoelectronic and photonic devices \cite{Jari2014}. Similar to graphene, monolayer TMDs are composed of a two-dimensional honeycomb lattice (Fig. \ref{Fig1}(a)) and can be isolated through mechanical exfoliation or grown using chemical vapor deposition and physical vapor transport techniques. As the material thickness is reduced to a single monolayer, TMDs transition from an indirect bandgap semiconductor to one with a direct gap at the two inequivalent $K$ and $K'$ momentum valleys located at the edges of the Brillouin zone, resulting in a thousand-fold increase in optical emission at visible wavelengths (Fig. \ref{Fig1}(b)) \cite{Mak2010,Splendiani2010}.

The unique combination of time-reversal symmetry, broken inversion symmetry, and strong spin-orbit splitting in monolayer TMDs leads to coupled spin and valley physics \cite{Xiao2012,Xu2014}. At the conduction and valence band edges, the orientation of the electronic spin is locked with the valley pseudospin degree of freedom, resulting in chiral optical selection rules: band-edge optical transitions at the $K$ valley are coupled to $\sigma+$ polarized light at normal incidence, while transitions in the $K'$ valley couple to $\sigma-$ polarized light (Fig. \ref{Fig1}(c)).

\begin{figure}[htbp!]
\centering
\includegraphics[width=0.5\linewidth]{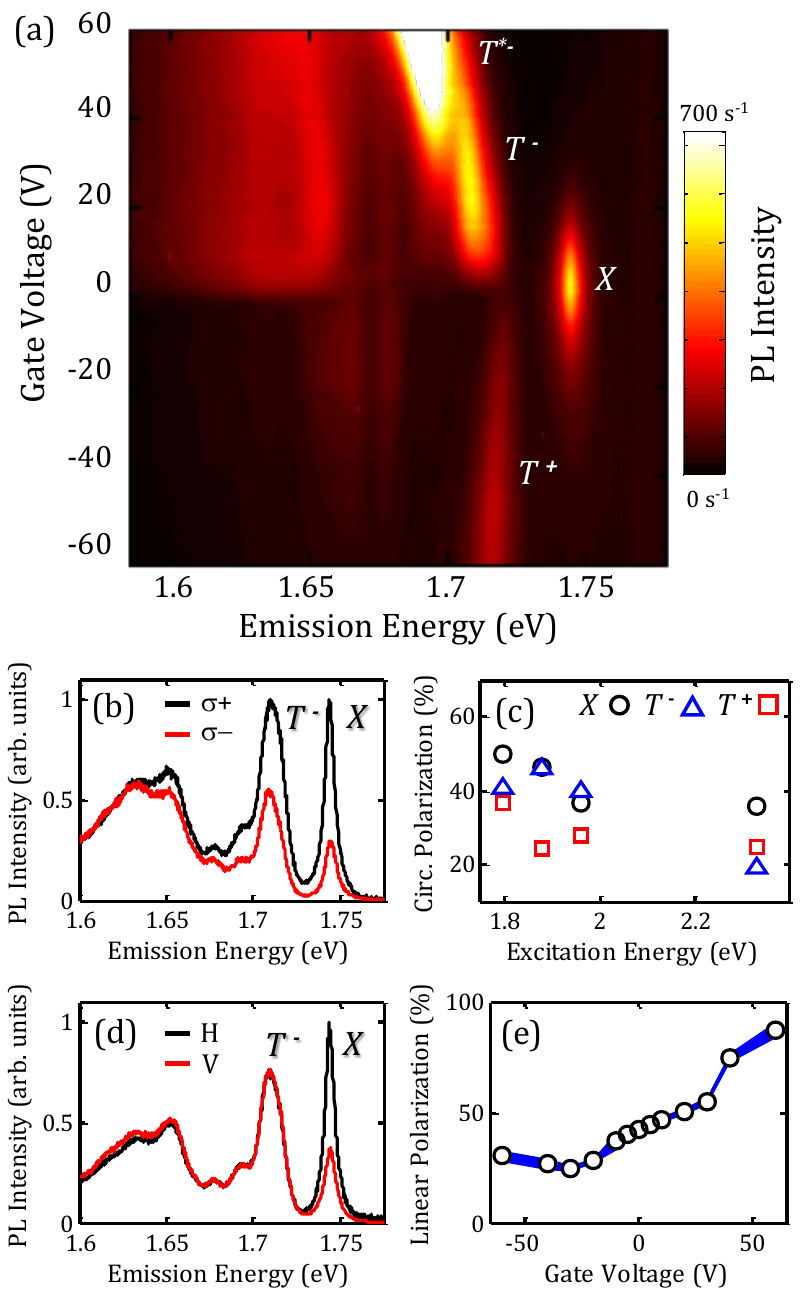}
\caption{(a) Photoluminescence intensity map of exfoliated monolayer WSe$_2$ as a function of electrostatic gate voltage and photon emission energy showing exciton (X), negative trion (T$^-$) and its fine structure (T$^{*-}$), and positive trion (T$^+$) transitions. (b) Polarization-resolved photoluminescence spectrum at +10 V gate voltage for $\sigma+$ (black curve) and $\sigma-$ (red curve) detection. The incident laser is $\sigma+$ polarized. (c) The degree of exciton and trion polarization decreases with increasing pump photo-excitation energy. (d) Polarization-resolved photoluminescence spectrum at +10 V gate voltage for horizontally (H, black curve) and vertically (V, red curve) polarized detection. The incident laser is horizontally polarized. (e) The degree of exciton linear polarization as a function of gate voltage. Data reproduced from \cite{Jones2013}.}
\label{Fig2}
\end{figure}

The optical spectra of monolayer TMDs feature pronounced peaks associated with exciton-type transitions, which inherit the dichroic optical selection rules \cite{Mak2012,Zeng2012,Cao2012}. The heavy electron and hole effective masses arising from the atomic $d$-orbitals, combined with reduced dielectric screening in two dimensional systems, leads to exceptionally strong Coulomb interactions and correlations between the charge carriers. Electron-hole pairs form tightly bound excitons with a $\sim 1$ nm Bohr radius and a $\sim 500$ meV binding energy--over an order of magnitude larger than conventional semiconductors \cite{Chernikov2014,Ugeda2014,He2014,Klots2014,Ye2014,Zhang2014,Wang2015,Zhu2015}. Through electrostatic gating, the exciton charge state can be modified, as illustrated in the photoluminescence intensity map in Fig. \ref{Fig2}(a) \cite{Jones2013}, which features neutral and charged excitons (trions) \cite{Mak2013,Ross2013,Berk2013}. These excitonic states exhibit rich many-body interaction effects, such as excitonic molecules (biexcitons) \cite{You2015,Shang2015,Sie2015} and strong coupling between excitons and trions \cite{Singh2014,Singh2015,Jones2015}.

The robust optical selection rules for the exciton and trion transitions are illustrated by the steady-state photoluminescence spectrum in Fig. \ref{Fig2}(b). After optical excitation of excitons in the $K$ valley using $\sigma+$ polarized light, emission primarily occurs from excitons in the same valley, indicating a high degree of valley polarization, defined as $\rho_c = \left(I_+ - I_-\right)/\left(I_+ + I_-\right)$, where $I_+$ ($I_-$) is the $\sigma+$ ($\sigma-$) polarized emission intensity (or horizontally (H) and vertically (V) polarized emission for linear excitation and detection). As the photo-excitation energy is increased further above the band edge, intervalley carrier scattering occurs with a higher probability and the degree of polarization decreases as shown in Fig \ref{Fig2}(c). Alternatively, one can excite the exciton transitions using linearly polarized light, shown in the photoluminescence spectra in Fig. \ref{Fig2}(d). The $\sigma+$ and $\sigma-$ components of the linearly polarized pump photons excite charge carriers in both the $K$ and $K'$ valleys. The electronic wavefunctions evolve with a fixed relative phase relationship during the hot-carrier relaxation, exciton formation, and emission processes. As a result, the emitted photons at the exciton energy are linearly polarized primarily along the pump photon polarization direction--a signature of optically induced valley coherence \cite{Jones2013,Wang2015}. Valley coherence is sensitive to the electronic wavefunction overlap and band alignment, which can be modified by an electrostatic field in gated samples (Fig. \ref{Fig2}(e)).

The high degree of valley polarization and valley coherence offers novel opportunities for manipulating the valley pseudospin degree of freedom. In order to leverage these unique physical properties of TMDs for valleytronic and optoelectronic applications, however, a clear understanding of the exciton radiative and non-radiative recombination, valley polarization, and valley coherence \textit{dynamics} is required. One can get a sense of the relative weight of radiative and non-radiative recombination by measuring the absolute quantum yield--that is, by comparing the number of photons emitted from the monolayer to the number of photo-excited excitons. As-grown monolayers have exhibited poor luminescence absolute quantum yield ranging from $<0.1\%$ to $6\%$, implying that ultrafast non-radiative decay channels compete with and can even dominate radiative recombination \cite{Mak2010,Rana2015_1}.

In order to gain a deeper understanding of the processes governing the recombination dynamics, time-resolved photoluminescence and ultrafast nonlinear optical spectroscopy techniques have been implemented. The exciton population decay dynamics typically exhibit multiple exponential components with time constants ranging from hundreds of femtoseconds \cite{Poellmann2015,Moody2015} to ten nanoseconds \cite{Amani2015}. At present, however, a consensus on the microscopic origins of the fast and slow time constants is still lacking. On the one hand, the strong exciton oscillator strength, resulting in up to $\sim20\%$ optical absorption for the exciton transition in a single monolayer \cite{Tonndorf2013}, implies efficient radiative recombination. Calculations of the radiative lifetime for excitons with zero center-of-mass momentum predict a sub-picosecond radiative decay \cite{Moody2015,Palummo2015}. On the other hand, the low quantum yield and large variance in the reported population lifetimes point towards a strong influence on the recombination dynamics from surface states, impurities, defects, and excitation conditions \cite{Rana2015_2,Amani2015}.  In the following sections, we review the experimental and theoretical findings that have led to these conclusions and discuss possible experiments to enhance our understanding of exciton recombination dynamics in monolayer TMDs.

\section{Population Recombination Dynamics}

Connecting the exciton decay dynamics to radiative and non-radiative processes can be ambiguous in time-resolved spectroscopy experiments due to the presence of defect and impurity states, which often appear in steady-state photoluminescence spectra as broad peaks red-shifted from the exciton resonance by $\sim100$ meV (see Fig. \ref{Fig2}). Interpretation of the exciton dynamics and comparison between different studies is also complicated by the presence of low-lying dark exciton states and bright edge states on the micron-sized monolayer flakes \cite{Gutierrez2013,Yin2014}. These states provide additional relaxation channels for charge carriers and may or may not contribute to exciton recombination depending on the excitation conditions and sample preparation. Such a scenario has stimulated numerous studies aiming to unambiguously separate intrinsic exciton radiative recombination from non-radiative decay processes.

\begin{figure*}[htbp!]
\centering
\includegraphics[width=0.9\columnwidth]{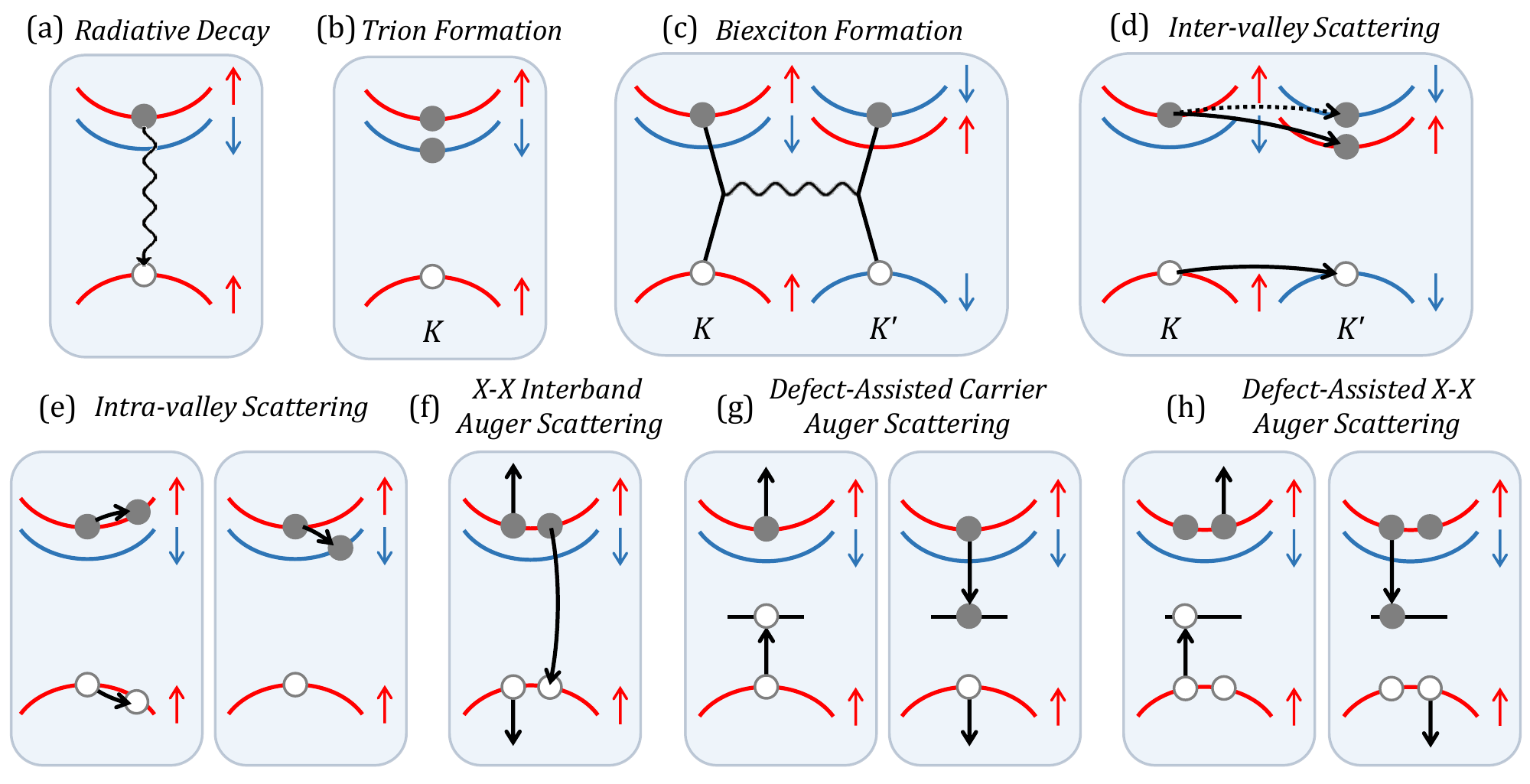}
\caption{Schematic illustration of the prevalent exciton radiative and non-radiative recombination processes in monolayer TMDs. The conduction and valence bands are labeled with the electronic spin orientations, shown in this case for WX$_2$. The filled and empty symbols represent an electron and hole, respectively. Radiative recombination can occur through the exciton transition (a). Alternatively, the exciton can bind with a free carrier to form a trion (b) or with another exciton to form a biexciton state (b), which subsequently emit photons red-shifted by the trion and biexciton binding energy, respectively. (d) Inter-valley scattering of either the electron or hole (or both) can occur. Only the case in which both the electron and hole scatter to a spin state with inverted orientation in the $K'$ valley results in an optically active bright exciton. (e) The center-of-mass momentum and phase of a bright exciton can be altered through intra-valley scattering of the electron and/or hole (left panel). A bright exciton becomes optically dark via a spin flip of the constituent electron (right panel). Non-radiative recombination can also occur through (f) interband exciton-exciton Auger scattering, (g) defect-assisted carrier-carrier Auger scattering, and (h) defect-assisted exciton-exciton Auger scattering.}
\label{Fig3}
\end{figure*}

The prevalent radiative and non-radiative relaxation channels in monolayer TMDs are illustrated by the energy diagrams in Fig. \ref{Fig3}. The lowest (highest) energy conduction (valence) band states at the $K/K'$ valleys are identified and labeled with the corresponding electronic spin orientation, shown in this case for WX$_2$. The electronic spin polarization of electrons in the lowest energy conduction band are expected to be opposite to that of holes in the upper valence band, in contrast to MoX$_2$ \cite{Liu2014}. In the majority of spectroscopy experiments, which use non-resonant optical excitation, hot carriers are excited high in the bands and then cool on a sub-picosecond timescale followed by exciton formation. The exciton-bound electron and hole can radiatively or non-radiatively recombine, or a combination of both, to give rise to multi-exponential decay dynamics.

\begin{figure}[htbp]
\centering
\includegraphics[width=0.65\linewidth]{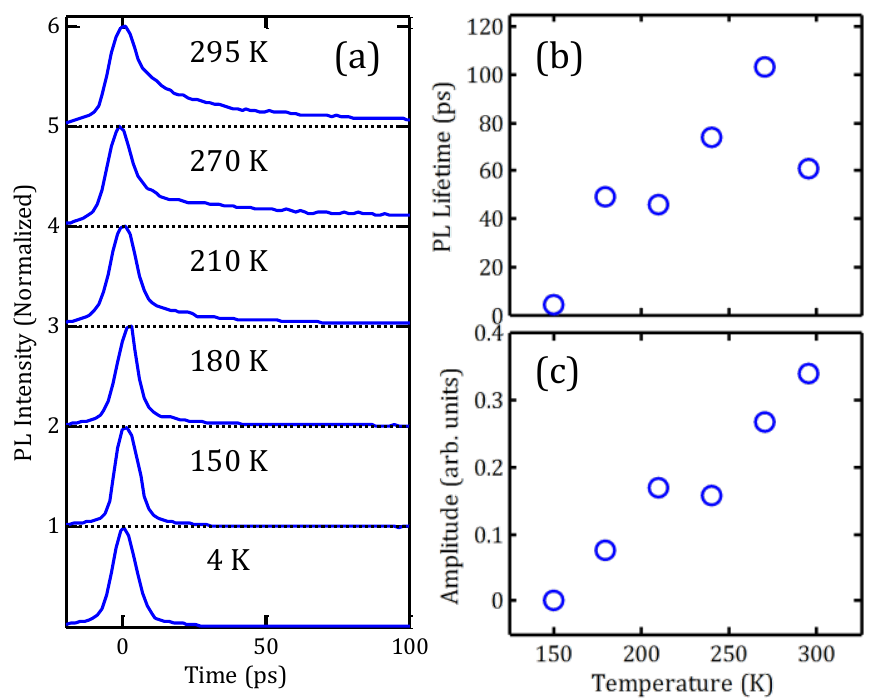}
\caption{(a) Temperature dependence of the photoluminescence decay dynamics of the lowest-energy exciton resonance in monolayer MoS$_2$ exfoliated onto a 300-nm-thick SiO$_2$/Si substrate. The sample is excited using a 532 nm laser. The time traces are offset vertically for clarity. (b) Above 150 K, a long-lived exponential decay component appears with a time constant that increases with temperature. (c) The amplitude of the long-lived component increases monotonically with temperature. Data reproduced from \cite{Korn2011}.}
\label{Fig4}
\end{figure}

The majority of time-resolved spectroscopy experiments have revealed two or three exponential decay time constants characterizing the recombination dynamics. Using time-resolved photoluminescence spectroscopy, Korn \textit{et al.} measure a biexponential response, shown in Fig. \ref{Fig4}(a) for monolayer MoS$_2$ exfoliated onto 300-nm thick SiO$_2$ on a silicon substrate \cite{Korn2011}. Below 150 K, the exponential population dynamics exhibit a single exponential response with a decay time that is resolution-limited at 5 ps. Above 150 K, a long-lived component appears and with a decay time increasing up to 100 ps at 270 K (Fig. \ref{Fig4}(b)). The amplitude of the long-lived component also increases monotonically with temperature (Fig. \ref{Fig4}(c)). This behavior is attributed to exciton-phonon intra-valley coupling of carriers to higher momentum states (left panel in Fig. \ref{Fig3}(e)), which changes the exciton center-of-mass momentum by scattering it to an optically inactive dark state outside of the light-cone, \textit{i.e.} the region in momentum space within which the emission of a photon can occur while energy and momentum conservation laws are obeyed. After scattering out of the light cone, the exciton center-of-mass momentum must be reduced through additional scattering events before the exciton can radiatively recombine. At elevated temperatures, exciton-phonon scattering occurs with a higher probability, resulting in a larger amplitude and slower decay time of the long-lived component.

First-principles calculations of the intrinsic exciton radiative lifetime lend additional credence to this interpretation. Palummo \textit{et al.} combine density functional theory (DFT) and the \textit{GW}-Bethe Salpeter equation method to compute the excitonic band structure, absorption spectrum, and wavefunction in monolayer TMDs \cite{Palummo2015}. Using Fermi's Golden rule, the radiative lifetime of excitons with zero center-of-mass momentum and at zero temperature is predicted to range from 190 fs to 240 fs depending on the constituent $M$ and $X$ atoms. Sub-picosecond radiative recombination has been reported in a recent ultrafast optical-pump and THz-probe spectroscopy study of monolayer WSe$_2$ \cite{Poellmann2015}. In this work, Poellmann \textit{et al.} measure a 150 fs lifetime component, which is attributed to radiative recombination of bright excitons within the light cone with near-zero center-of-mass momentum. A rate equation analysis taking into account both bright and dark exciton transitions yields consistent results between the fast radiative decay of low-momentum excitons and the overall low quantum yield from their sample ($\sim10\%$).

In the majority of ultrafast spectroscopy experiments, however, the average response of a thermal distribution of excitons in momentum space is probed. Excitons are expected to thermalize on a timescale faster than the radiative lifetime, which can be attributed to exciton-phonon inelastic scattering processes that change the in-plane exciton momentum \cite{Shi2013}. A thermalized distribution of excitons can be described using Boltzmann statistics according to $N = exp(-E_k/k_bT)$, where $E_k$ is the exciton energy, $k_B$ is the Boltzmann constant, and $T$ is the sample temperature. When considering radiative recombination, the response of the distribution of excitons must be taken into account, leading to an \textit{effective} radiative lifetime. Palummo \textit{et al.} predict that the effective lifetime increases linearly with temperature at a rate of $\sim1-10$ ps/K and ranges from 1-10 ps at 4 K to 1-5 ns at room temperature \cite{Palummo2015}. These values are comparable to the low-temperature lifetime and linear increase with temperature reported by Korn \textit{et al.} shown in Fig. \ref{Fig4}(b), although an order-of-magnitude discrepancy exists at room temperature. A fast 4.5 ps ($\leq3$ ps) exciton lifetime at 4 K is also reported in exfoliated MoS$_2$ (MoSe$_2$) using time-resolved photoluminescence \cite{Glazov2015,Wang2015_10}. At elevated temperatures, tri-exponential exciton recombination dynamics have been identified in suspended MoS$_2$ using transient absorption spectroscopy \cite{Shi2013}. Shi \textit{et al.} report time constants of 2 ps, 75 ps, and 850 ps; the long time component is in excellent agreement with the above calculations for an average thermalized distribution of excitons at room temperature.

At cryogenic temperatures, ultrafast quenching of exciton emission on a sub-picosecond to few-picosecond timescale has been attributed to several competing factors. The fast and intermediate lifetimes measured by Shi \textit{et al.} might be related to the presence of defects, impurities, edge states, and substrate effects, discussed in more detail below. Additionally, low-lying optically dark exciton states can influence the fast and slow recombination dynamics \cite{Zhang2015_1,Palummo2015}. Emission from the optically bright exciton transition can be quenched on a picosecond timescale through inter- and intra-valley scattering and spin flip of the electron to form a dark exciton, illustrated by the processes in Fig. \ref{Fig3}(d) and the right panel of Fig. \ref{Fig3}(e) \cite{Wang2013,Mai2014,Zhu2014}. At elevated temperatures comparable to or larger than the conduction band splitting, both the bright state and low-lying dark state are occupied according to Boltzmann statistics. Zhang \textit{et al.} provide evidence that both exciton states are populated in exfoliated WSe$_2$ above 80 K, resulting in a long lifetime component attributed to both fast non-radiative recombination and slower radiative recombination on the order of one nanosecond \cite{Zhang2015_1}. However, below $80$ K, a thermal equilibrium is not established between these states, and the dark states serve as a fast non-radiative relaxation channel that competes with radiative recombination on a $\sim10$ ps timescale (limited by the instrument response function in this particular study) \cite{Zhang2015_1}; however, the nature of and characteristic timescale for bright-to-dark state conversion is likely sensitive to the material composition. For example, the inverted spin polarization and smaller conduction band splitting in MoX$_2$ compared to WX$_2$ (2-3 meV compared to tens of meV, respectively) may result in a thermal equilibrium between bright and dark states even at cryogenic temperatures, potentially eliminating this contribution to the quenching of exciton emission in MoSe$_2$ and MoS$_2$.

\begin{figure}[htbp]
\centering
\includegraphics[width=0.5\linewidth]{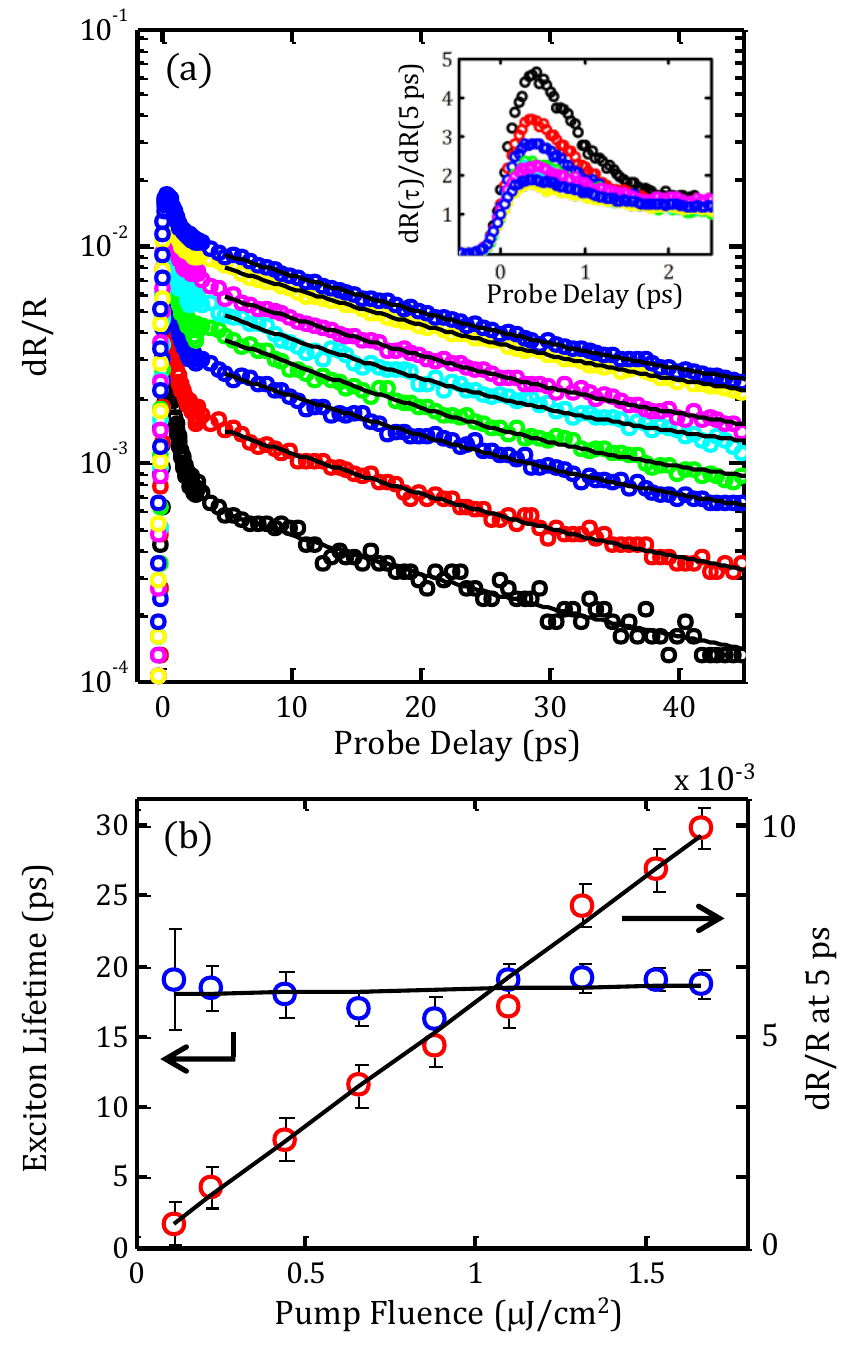}
\caption{Room temperature differential reflectance spectroscopy of monolayer WSe$_2$ exfoliated onto a silicon substrate with a 280-nm thick SiO$_2$ layer. The differential reflection dynamics measured using a 750 nm probe and 405 nm pump are shown in (a) for pump fluence increasing from 0.11 to 1.66 $\mu$J/cm$^2$ (bottom to top). Inset: signal dynamics at early probe delays. (b) The long-lived exciton lifetime component (blue circles, left axis) and the signal intensity (red circles, right axis) as a function of pump fluence. Data reproduced from \cite{Cui2014}.}
\label{Fig5}
\end{figure}

The ultrafast recombination dynamics have also been attributed to exciton and carrier Auger-type processes. The two-dimensional nature of TMDs and the heavy effective masses of the carriers enhance the Coulomb interactions and correlations between charge carriers, which results in exceptionally strong exciton-exciton and carrier-carrier scattering in these materials \cite{Wang2015_5}. Scattering between two excitons can induce non-radiative interband recombination of one exciton and subsequent scattering of the other, conserving energy and momentum as illustrated in Fig. \ref{Fig3}(f). Exciton-exciton annihilation has been verified experimentally through exciton density dependent measurements of the recombination dynamics at room temperature in exfoliated MoS$_2$, measured using ultrafast transient absorption spectroscopy \cite{Sun2014}. Through a rate equation analysis that includes a decay channel varying quadratically with the exciton density, Sun \textit{et al.} extract an exciton-exciton annihilation rate of 0.04 cm$^2$/s, implying an effective exciton lifetime of $\sim10$ ps at an exciton density of $10^{12}$ cm$^{-2}$ (exciton-exciton separation of 10 nm). At comparable excitation densities in MoSe$_2$, an order-of-magnitude larger exciton-exciton annihilation rate of 0.33 cm$^2$/s is reported \cite{Kumar2014}. At these exciton densities, non-radiative recombination through biexciton formation also provides an efficient relaxation channel with a quadratic dependence of the signal on the pump fluence \cite{Sie2015,You2015}.

In the low-density regime ($\leq 10^{12}$ cm$^{-2}$), extrinsic factors can influence exciton recombination dynamics and can even dominate over exciton-exciton interband annihilation. It is reasonable to speculate that the two-dimensional nature of atomically thin TMDs leads to a strong influence on the optical and electronic properties from point defects such as $M$ and $X$ vacancies, interstitial sites, impurity atoms, and grain boundaries \cite{Zande2013,Zhou2013}. Impurity atoms can introduce free carriers into the system, resulting in non-radiative decay of the bright exciton population through an exciton-to-trion conversion process via charge-carrier capture by the exciton on a picosecond timescale (Fig. \ref{Fig3}(b)) \cite{Singh2015}.

Defect and impurity states also serve as a center for efficient non-radiative recombination provided the final state wavefunction strongly overlaps with the Bloch states of the valence and conduction bands. In a simple picture, an electron (hole) can scatter off a hole (electron), consequently being captured into a defect or impurity state while the hole (electron) scatters to a higher energy state to conserve energy, as illustrated in Fig. \ref{Fig3}(g). The capture time into defect states via carrier-carrier Auger scattering is predicted to vary from 0.5-3 ps and is independent of the pump fluence for exciton densities ranging from $10^{11}-10^{12}$ cm$^{-2}$ \cite{Wang2015_5}. At excitation densities $\geq 10^{12}$ cm$^{-2}$, a combination of linear and quadratic pump fluence dependence of the carrier capture rates into defect states is expected; in contrast to exciton-exciton interband annihilation, this nonlinear behavior arises from the capture of an electron or hole from one exciton and concurrent scattering of a hole or electron in the other (see Fig. \ref{Fig3}(h)). While the former does not require the presence of defect and impurity states, the probability of the latter increases as the defect energy within the bandgap approaches the conduction (valence) band for electron (hole) capture \cite{Wang2015_5}.

Density independent Auger scattering may be reflected in the room temperature differential reflectance spectra shown in Fig. \ref{Fig5}, measured from monolayer WSe$_2$ exfoliated onto a 280-nm thick SiO$_2$ layer on a silicon substrate \cite{Cui2014}. In this study, Cui \textit{et al.} report a constant $\sim18$ ps exciton lifetime for excitation densities up to $10^{11}$ cm$^{-2}$, which is consistent with the density-independent carrier capture rates predicted by Wang \textit{et al.} \cite{Cui2014,Wang2015_5}. The short exciton lifetime in Fig. \ref{Fig5}(b) is also tentatively attributed to phase-space filling, Coulomb screening, and bandgap renormalization effects. Mouri \textit{et al.} report similar results for WSe$_2$ on quartz for exciton densities near $10^{11}$ cm$^{-2}$; however, as the density is further decreased to $10^9$ cm$^{-2}$, the lifetime increases to $\sim4$ ns, which suggests minimal contributions from defect-assisted non-radiative recombination in this study. This highly nonlinear behavior is attributed to diffusion-assisted exciton-exciton annihilation at a rate of 0.35 cm$^2$/s and a diffusion length up to 1.8 $\mu$m \cite{Mouri2014}. The large variation in reported annihilation rates and exciton lifetimes for a given excitation density might be attributed to a strong influence from the substrate. For example, the exciton-exciton annihilation rate decreases by more than a factor of two when transferring from a supported substrate to a suspended monolayer in WS$_2$ and MoS$_2$, which can significantly affect the quantum yield \cite{Yu2015}.

Ultrafast non-radiative decay of a thermalized distribution of excitons, as observed in the above-mentioned studies, is consistent with the low quantum yield measured through steady-state photoluminescence spectroscopy. The role of defect and surface traps on exciton recombination has been explored through lifetime measurements as a function of the TMD layer thickness. Using ultrafast differential transmission spectroscopy, Wang \textit{et al.} demonstrate that the exciton lifetime increases from a few tens of picoseconds for a single monolayer to a nanosecond for 10 layers in exfoliated MoS$_2$ at room temperature. This behavior is attributed to a short defect-assisted recombination lifetime for the surface layers in a few-layer sample and a long recombination lifetime for all the inner layers. The effective lifetime is estimated by taking into account the probability of electron and hole occupation in each layer using the effective mass approximation. The short exciton lifetime in the monolayer is attributed to fast defect-assisted carrier recombination via Auger scattering, which decreases in probability as the sample thickness is increased \cite{Rana2015_2}. Similar surface-recombination dynamics have been observed in more conventional semiconductor nanostructures, which have benefitted from passivation schemes that reduce the number of available non-radiative recombination sites \cite{Yablo1986,Dan2011}. In monolayer TMDs, specifically exfoliated MoS$_2$, treatments with a non-oxidizing organic superacid, bis(trifluoromethane) sulfonimide (TFSI), increase the exciton lifetime measured at room temperature from 100 ps to more than 10 ns and enhance the quantum yield by more than two orders of magnitude--up to $95\%$. \cite{Amani2015}. This air-stable, solution-based passivation process points toward a systematic method for minimizing contributions to exciton recombination dynamics from defects and impurities.

\section{Coherent Dynamics}

\begin{figure}[htbp!]
\centering
\includegraphics[width=0.7\linewidth]{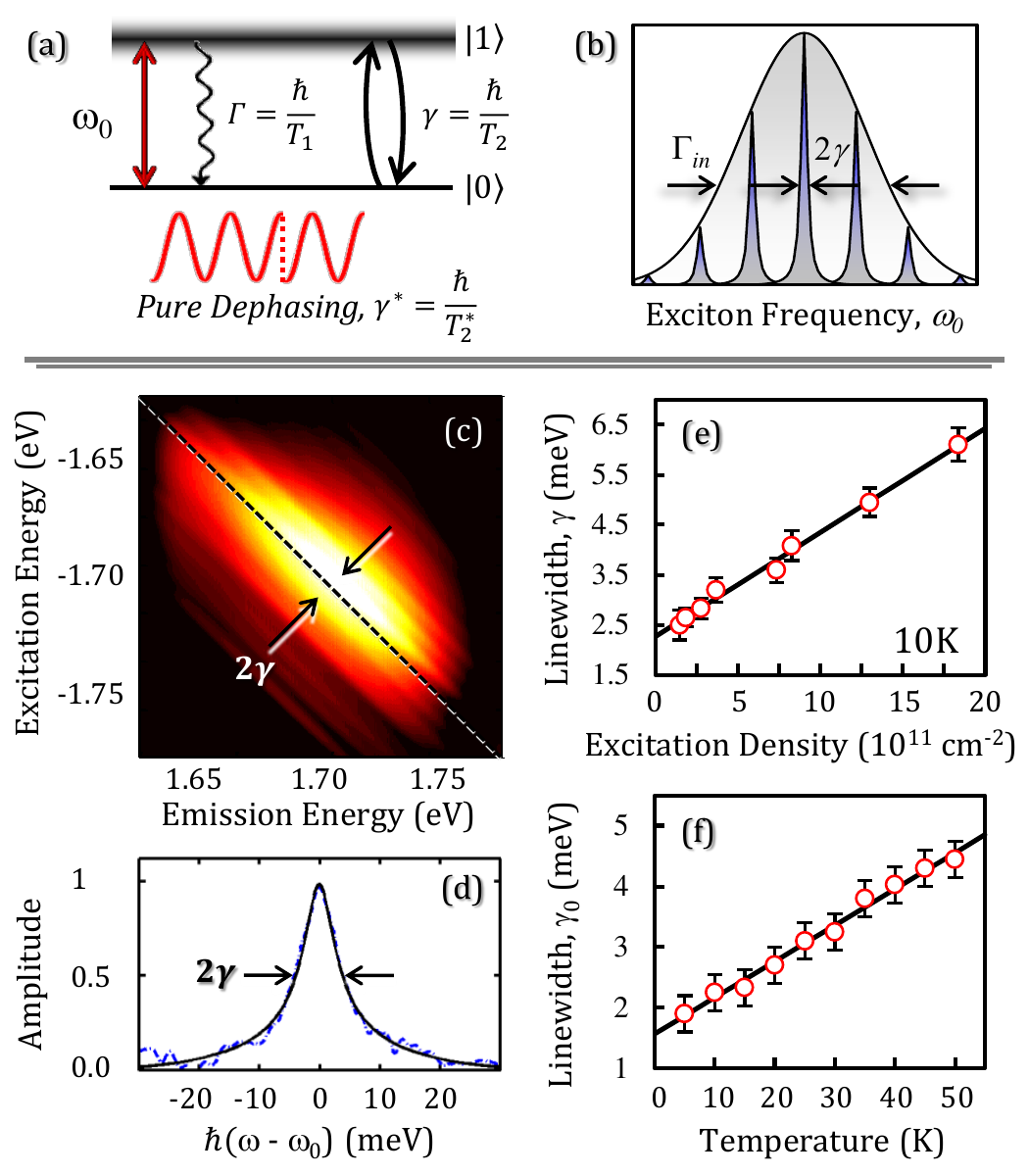}
\caption{(a) The coherent quantum dynamics of an exciton with resonance frequency $\omega_0$ are characterized by the recombination rate $\Gamma$ (lifetime $T_1 = \hbar/\Gamma$) and homogeneous linewidth $\gamma$ (coherence time $T_2 = \hbar/\gamma$). The two are related through $\gamma = \Gamma/2 + \gamma^*$, where $\gamma^*$ is the pure dephasing rate describing elastic processes that interrupt the phase coherence without affecting the excited-state occupation. (b) Inhomogeneous broadening ($\Gamma_{in}$) due to a varying local potential results in a distribution of exciton transition frequencies. (c) Two-dimensional coherent spectrum of the exciton transition at 5 K. The linewidth along the diagonal dashed line provides a measure of the inhomogeneous linewidth, whereas the half-width at half-maximum along the cross-diagonal line provides the homogeneous linewidth. (d) A cross-diagonal slice taken at the maximum amplitude of the spectrum in (c) is fit with a square root of Lorentzian function to yield $\gamma = 2.7$ meV at an excitation density of 10$^{11}$ cm$^{-2}$. (e) The homogeneous linewidth increases linearly with exciton density up to $\sim10^{12}$ cm$^{-2}$, shown for 10 K. (f) The extrapolated zero-density homogeneous linewidth as a function of temperature. Data reproduced from \cite{Moody2015}.}
\label{Fig6}
\end{figure}

While time-resolved photoluminescence and pump-probe techniques provide essential information regarding exciton radiative and non-radiative recombination ($T_1$ dynamics), they do not provide any details of the exciton coherent dynamics. The exciton coherence time ($T_2$)--which reflects the timescale during which a superposition of the crystal ground and excited exciton states evolves with a fixed phase relationship--is a fundamental parameter of light-matter interaction in semiconductors (see Fig. \ref{Fig6}(a)). In principle, $T_1$ and $T_2$ reflect the fundamental timescales for coherent opto-electronic, photonic, and quantum information applications. The coherence time can be probed in either the time or frequency domains; however impurities and defects give rise to local potentials that shift the exciton transition, resulting in an inhomogeneous distribution of exciton frequencies. In steady-state photoluminescence spectra, the exciton linewidth is a convolution of the intrinsic homogeneous linewidth (inversely proportional to $T_2$) and the inhomogeneous linewidth, which typically dominates the optical spectra as shown in Fig. \ref{Fig6}(b).

To extract exciton homogeneous broadening from inhomogeneous broadening in TMDs, optical two-dimensional coherent spectroscopy (2DCS) has been employed, which is a three-pulse four-wave mixing (photon echo) technique \cite{Moody2015}. In this study, a series of 100-fs laser pulses with phase-stabilized, variable delays coherently interact with a monolayer WSe$_2$ sample grown via chemical vapor deposition on a sapphire substrate. The coherent light-matter interaction generates a third-order polarization that is radiated as a photon echo, which is detected through heterodyne spectral interferometry. A two-dimensional Fourier transform of the four-wave mixing signal with respect to the varied delays generates a two-dimensional spectrum that correlates the excitation and emission energies of the system, shown in Fig. \ref{Fig6}(c). The spectrum features a single peak centered at the exciton absorption energy. The linewidth along the diagonal dashed line reflects the amount of inhomogeneous broadening in the material, whereas the half-width at half-maximum of the cross-diagonal lineshape provides a measure of the homogeneous linewidth ($\gamma$), as shown in Fig. \ref{Fig6}(d) \cite{Siemens2010}.

The narrow homogeneous linewidth, on the order of a few meV, compared to the $\sim50$ meV inhomogeneous linewidth, confirms the presence of varying local potentials via defects and impurities in this particular study. Weak localization leading to inhomogeneous broadening of the exciton resonance is also observed in exfoliated monolayer samples, which results in the appearance of an exciton ``mobility edge'' separating ``localized'' and ``delocalized'' states within the exciton transition \cite{Singh2015}. While the exciton lifetime is independent of the excitation density below $10^{12}$ cm$^{-2}$ as discussed in the previous section, the homogeneous linewidth increases linearly with density as shown in Fig. \ref{Fig6}(e).
This behavior is reminiscent of exciton excitation-induced dephasing in conventional semiconductors that arises from elastic scattering of excitons \cite{Boldt1985}. Interestingly, the amount of exciton-exciton interaction broadening in TMDs is nearly an order of magnitude larger than traditional semiconductor nanostructures, which is attributed to reduced dielectric screening of the Coulomb interaction in TMDs.

\begin{figure}[htbp!]
\centering
\includegraphics[width=0.5\linewidth]{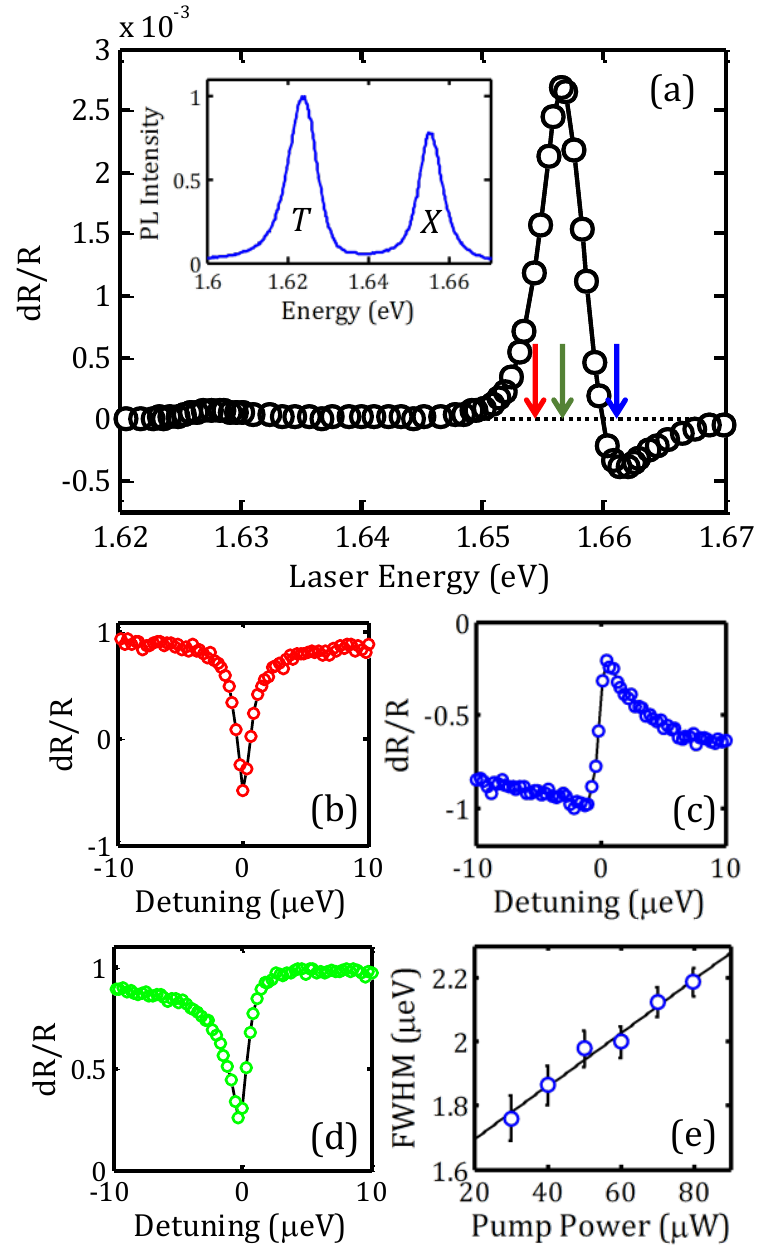}
\caption{(a) Degenerate differential reflectance (dR/R) spectrum from exfoliated monolayer MoSe$_2$ taken at 30 K featuring the exciton and trion resonances near 1.66 and 1.63 eV, respectively. The peak assignments are consistent with the low temperature photoluminescence spectrum (inset). The colored arrows indicate the spectral position of the pump in the non-degenerate spectra. (b)-(d) Non-degenerate differential reflectance spectra as a function of pump energy tuned throughout the exciton resonance for cross-linearly polarized pump and probe. The linewidth extracted from the real and imaginary components of a complex Lorentzian fit function are shown in (e). The linewidth full-width at half-maximum (FWHM) increases linearly with the pump power for 20 $\mu$W probe power. Data reproduced from \cite{Schaibley2015}.}
\label{Fig7}
\end{figure}

The role of acoustic phonons in exciton dephasing is reflected in the temperature dependence of the homogeneous linewidth. At each temperature, the excitation density dependence of the linewidth is measured. The extrapolated zero-density linewidths are shown in Fig. \ref{Fig6}(f) for temperatures up to 50 K. The residual homogeneous linewidth extrapolated to zero density and zero temperature is $\gamma=1.6$ meV. This value is limited only by the exciton recombination lifetime in this study (a few-hundred femtoseconds), obtained from a variation of 2D spectroscopy that is equivalent to optical pump-probe techniques. This is an interesting result, since it implies that the fast radiative and non-radiative recombination dynamics leading to the sub-picosecond exciton lifetime do not introduce additional pure dephasing in this sample, \textit{i.e.} $\gamma^* = 0$.

Additional insight into the coherent quantum dynamics of excitonic transitions in monolayer TMDs is provided through ultra high-resolution differential reflectance (DR) spectroscopy. In this study, the authors perform a two-color continuous-wave pump-probe experiment on monolayer MoSe$_2$ exfoliated onto an SiO$_2$/Si substrate at 30 K \cite{Schaibley2015}. A pump laser is resonant with the exciton resonance while a probe laser is scanned through zero pump-probe detuning. Through modulation of the pump and probe amplitudes and lock-in detection referenced to the difference in modulation frequencies, the technique is sensitive to ultra-narrow resonances corresponding to ultra-long decay dynamics. A degenerate DR spectrum obtained by scanning the pump and probe frequencies for zero pump-probe detuning is shown in Fig. \ref{Fig7}(a), which is used to characterize the sample. Comparing the DR nonlinear optical response to the photoluminescence spectrum (inset in Fig. \ref{Fig7}(a)), the two resonances are attributed to the exciton and trion transitions at 1.655 and 1.625 eV, respectively.

In contrast to the few-meV linewidth of the exciton resonance in the degenerate DR spectrum, the high-resolution two-color DR measurements feature $\mu$eV-wide resonances superimposed onto an meV-wide background that closely resembles the degenerate DR spectrum. Two-color DR spectra are shown in Fig. \ref{Fig7}(b)-\ref{Fig7}(d) for cross-linearly polarized pump and probe fields at three different pump energies. Each spectrum is fit with a complex Lorentzian function to extract a linewidth. A power dependence of the linewidth is shown in Fig. \ref{Fig7}(e) for the pump tuned to the peak of the exciton resonance in the degenerate DR spectrum (corresponding to the lineshape in Fig. \ref{Fig7}(d)). Extrapolating to zero pump power reveals a linewidth of $\sim1.5$ $\mu$eV.

Both spectral-hole burning and coherent population oscillation nonlinearities contribute to the measured linewidths in Fig. \ref{Fig7}. While the few-meV background is attributed to spectral hole burning, which provides a measure of homogeneous and inhomogeneous broadening, the $\mu$eV width is associated with coherent population oscillations that reflect a long-lived (nanosecond) component. Using a model based on the optical Bloch equations for a 5-level system, the authors attribute the long lifetime to a combination of both bright and dark exciton transitions that includes non-radiative decay to a long-lived state as well as inter-valley scattering. The pump fluence dependence of the linewidth shown in Fig. \ref{Fig7}(e) points to the influence of interaction effects such as exciton-exciton annihilation. Using co-circular polarization of the pump and probe fields, an additional sub-$\mu$eV component corresponding to nearly a 10 ns lifetime appears, indicating the presence of an additional long-lived state. Understanding the origin of this state might help explain the fast exciton recombination and low quantum yield at low temperature and the increase in exciton lifetime at elevated temperature in some experiments.

\section{Valley Polarization and Valley Coherence Dynamics}

The bright exciton spin and valley relaxation dynamics can be modeled within a pseudospin formalism (see \cite{Glazov2015} for a review). The pseudospin (\textbf{S}) components describe the orientation of the microscopic exciton dipole moment, \textit{i.e.} $S_z$ gives the degree to which excitons remain in their initial $K/K'$ valley (valley polarization), whereas $S_{x,y}$ give the degree of linear polarization corresponding to a coherent superposition of excitons in both $K$ and $K'$ valleys (valley coherence). Enhanced Coulomb interactions in monolayer TMDs lead to exciton spin and valley depolarization through the long-range exchange interaction. The pseudospin formalism readily takes into account intra- and inter-valley exchange effects \cite{Yu2013,Glazov2014,Yu2014}, the latter of which acts as an effective magnetic field that induces coupling between the $K$ and $K'$ valleys.

\begin{figure}[htbp]
\centering
\includegraphics[width=0.5\linewidth]{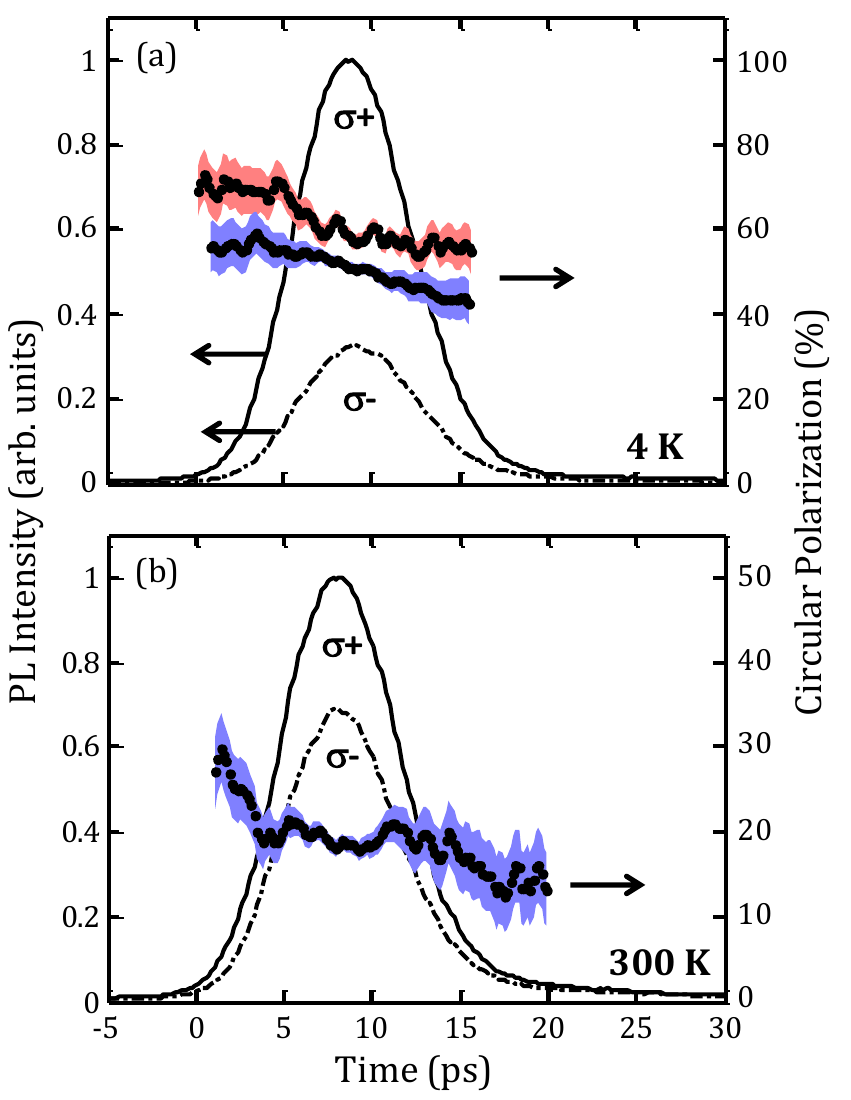}
\caption{Circularly polarized time-resolved photoluminescence intensity of the exciton resonance from exfoliated monolayer MoS$_2$ at (a) 4 K and (b) 300 K (left axes). The excitation laser is $\sigma+$ circularly polarized. Right axes: Degree of circular polarization during the exciton emission. Lowering the laser power by two orders of magnitude increases the degree of polarization, indicated by the red data points in (a). Data reproduced from \cite{Lagarde2014}.}
\label{Fig8}
\end{figure}

This formalism has been successfully applied to model the exciton spin and valley depolarization dynamics measured using time-resolved photoluminescence \cite{Glazov2014} and Kerr rotation spectroscopies \cite{Zhu2014}. The role that the long-range exchange interaction plays in inter-valley depolarization has also been identified through a combination of time-resolved and polarization-resolved photoluminescence \cite{Lagarde2014,Yan2015} and transient absorption \cite{Mai2014_2} spectroscopy studies. Figure \ref{Fig8} shows polarization- and time-resolved photoluminescence from monolayer MoS$_2$ exfoliated onto a 90-nm-thick SiO$_2$/Si substrate at two different temperatures (left axes) \cite{Lagarde2014}. Non-resonant $\sigma+$ polarized excitation is used and both $\sigma+$ and $\sigma-$ components (solid and dashed lines, respectively) of the emission are detected at the lowest energy exciton resonance. Exciton recombination occurs within $4$ ps at both 4 K and 300 K.

Interestingly, the degree of circular polarization is nearly constant for the entire duration of emission (blue data points in Figs. \ref{Fig8}(a) and \ref{Fig8}(b), right axes). As the excitation fluence is reduced by two orders of magnitude, the degree of circular polarization increases from $\sim50\%$ to $\sim60\%$ at 4 K. From the measured exciton lifetime, which includes contributions from both radiative and non-radiative recombination, and assuming an initial valley polarization of $100\%$, the extracted valley lifetime is $7$ ps. Similar timescales are also observed in exfoliated WSe$_2$ \cite{Wang2014_5}. Fast valley depolarization is consistent with the theoretical estimates from the pseudospin formalism \cite{Glazov2015}. The exciton spin and valley depolarization time has also been predicted to decrease with increasing temperature due to the long-range exchange interaction, which agrees well with Kerr rotation experiments \cite{Zhu2014}. Despite fast spin and valley relaxation, the degree of circular polarization as measured in steady-state photoluminescence experiments can be near $100\%$ owing to exciton recombination typically occurring on the same ultrafast timescale. Using circularly polarized transient absorption spectroscopy, a longer valley lifetime component on the order of $100$ ps has been measured, which is attributed to two processes: direct scattering of the exciton from the $K$ to $K'$ valley via a spin flip-flop interaction, and scattering of the exciton through a spin-degenerate $\Gamma$ valley \cite{Mai2014}.

The electron-hole exchange interaction also plays an important role in exciton valley coherence. In steady-state photoluminescence spectra, valley coherence is identified through the observation of linearly polarized emission oriented parallel to the linearly polarized excitation field \cite{Jones2013,Wang2015}. The $\sigma+$ and $\sigma-$ polarized components of the excitation field generate hot carriers in both the $K$ and $K'$ bands, which subsequently cool to form a thermal distribution of excitons that eventually recombines. The initial phase relationship of the circuarly polarized components of the excitation field is maintained during this process, resulting in linearly polarized emission. In these initial studies, a degree of linear polarization close to $50\%$ is reported, which decreases as the excitation energy increases above the exciton resonance energy. Such a behavior points to contributions from the electron-hole exchange interaction on the valley coherence dynamics. To probe these effects, the valley coherence time has been measured using two-dimensional coherent spectroscopy \cite{Hao2016}. In this work, the authors resonantly generate a valley coherence in WSe$_2$ using a series of circularly polarized laser pulses, analogous to a stimulated Raman-type process. The non-radiative valley coherence is converted to an optical coherence in the $K'$ valley that is detected through spectral interferometry. The delay between the laser pulses is varied to map out the valley coherence dynamics, which exponentially decay on a sub-picosecond timescale. A model based on the Maialle-Silva-Sham mechanism reproduces the measured valley dynamics \cite{Maialle1993,Vinattieri1994}, highlighting the role of both exciton population recombination and electron-hole exchange in exciton valley decoherence.

\section{Outlook}

\begin{table}[htbp!]
\centering
\caption{\bf Exciton Recombination Lifetime and Coherence Time in Monolayer TMDs}
\begin{adjustbox}{max width=\textwidth}
\begin{tabular}{ccc}
\hline
Material (Temp.) & Recombination ($T_1$) and Coherence ($T_2$) Times & Reference \\
\hline
WSe$_2$ (Various) & $T_{1,\textrm{fast}} = 210$ fs (10 K); $T_{1,\textrm{slow}} = 17$ ps (10 K); $T_2 = 150-410$ fs (50-0 K) & Moody \textit{et al.} \cite{Moody2015} \\
\hline
MoS$_2$ (300 K) & $T_{1} = 300$ ps (untreated); $T_{1} = 10.8$ ns (passivated) & Amani \textit{et al.} \cite{Amani2015} \\
\hline
MoS$_2$ (300 K) & $T_{1} = 50$ ps & Wang \textit{et al.} \cite{Rana2015_2} \\
\hline
MoS$_2$ (Various) & $T_{1,\textrm{fast}} = 5$ ps (5 K); $T_{1,\textrm{slow}} \approx 100$ ps (270 K) & Korn \textit{et al.} \cite{Korn2011} \\
\hline
MoS$_2$ (300 K, susp.) & $T_{1,\textrm{fast}} = 2.6$ ps; $T_{1,\textrm{mid}}=  74$ ps; $T_{1,\textrm{slow}}=  850$ ps & Shi \textit{et al.} \cite{Shi2013} \\
\hline
MoS$_2$ (300 K, supp.) & $T_{1,\textrm{fast}} = 3.3$ ps; $T_{1,\textrm{mid}}=  55$ ps; $T_{1,\textrm{slow}}=  469$ ps & Shi \textit{et al.} \cite{Shi2013} \\
\hline
MoSe$_2$ (4 K) & $T_{1} \leq 3$ ps & Wang \textit{et al.} \cite{Wang2015_10} \\
\hline
WSe$_2$ (Various) & $T_{1,\textrm{fast}} \leq 10$ ps ($\leq80$ K); $T_{1,\textrm{slow}} \approx 1$ ns (290 K) & Zhang \textit{et al.} \cite{Zhang2015_1} \\
\hline
WSe$_2$ (300 K) & $T_{1} = 18$ ps & Cui \textit{et al.} \cite{Cui2014} \\
\hline
MoS$_2$ (300 K) & $T_{1} = 19$ ps (22 $\mu$J/cm$^2$); $T_{1} \approx 360$ ps (3 $\mu$J/cm$^2$) & Sun \textit{et al.} \cite{Sun2014} \\
\hline
WSe$_2$ (300 K) & $T_{1} \leq 100$ ps (12 $\mu$J/cm$^2$); $T_{1} \approx 4$ ns (0.006 $\mu$J/cm$^2$) & Mouri \textit{et al.} \cite{Mouri2014} \\
\hline
MoSe$_2$ (30 K) & $T_{1,\textrm{fast}} = 1.7$ ns; $T_{1,\textrm{slow}} \approx 6$ ns & Schaibley \textit{et al.} \cite{Schaibley2015} \\
\hline
MoS$_2$ (4 K) & $T_{1} = 4$ ps & Lagarde \textit{et al.} \cite{Lagarde2014} \\
\hline
WSe$_2$ (4 K) & $T_{1,\textrm{fast}} \leq 4$ ps; $T_{1,\textrm{slow}} = 33$ ps & Wang \textit{et al.} \cite{Wang2014_5} \\
\hline
\end{tabular}
\end{adjustbox}
  \label{Table1}
\end{table}

The understanding of exciton recombination and decoherence dynamics has rapidly progressed since the seminal works identifying monolayer TMDs \cite{Mak2010,Splendiani2010}. Although research of monolayer TMDs only began in 2010, the breadth of experimental and theoretical studies have elucidated many important processes that govern exciton dynamics. Specifically, exciton thermalization has been identified as one leading factor dictating the recombination timescale. At room temperature, a thermal distribution of excitons in momentum space leads to an effective radiative lifetime on the order of 100 ps to 1 ns. At cryogenic temperatures, the extent of the distribution is reduced, resulting in a sub- to few-picosecond effective lifetime. Non-radiative recombination through Auger-type scattering and defect-assisted relaxation channels compete on a similar timescale. These contributions can be minimized using low pump fluence and surface-state passivation treatments, respectively, which extend the effective radiative lifetime to nearly 11 ns and enhance the quantum yield to $\sim95\%$ at room temperature. The presence of low-lying dark states appearing through intra- and inter-valley spin scattering can also lead to an additional fast relaxation channel at low temperature. In Table \ref{Table1}, we provide an overview of the exciton recombination and decoherence timescales reported in the literature to date.

Minimizing non-radiative recombination in monolayer TMDs is a highly sought goal that would enable novel high-performance opto-electronics and photonics applications. Future studies of exciton dynamics in monolayer TMDs may yield further insight by combining several of the techniques and control knobs discussed in this review into a single experiment. For example, time-resolved photoluminescence and Kerr rotation spectroscopy experiments on passivated samples, performed at various temperatures, may reveal longer exciton lifetimes that are compatible with electronically accessible ($\geq$ nanosecond) time scales. Combined with coherent nonlinear spectroscopy experiments, the full exciton quantum dynamics including population recombination and decoherence times could be characterized. Novel quantum phenomena and optoelectronic properties might be revealed by applying similar techniques to study other exciton-type states including charged and multi-excitons, localized quantum dot-like excitons \cite{Sriv2015,Chak2015,Koperski2015,He2015}, indirect excitons in bilayers \cite{Jones2014,Rivera2015} and heterostructures \cite{Chen2013,Britnell2013}, and exciton-polaritons \cite{Liu2014_2}.

\section*{Funding Information}

\textbf{J.S. and X.X. are supported by the Department of Energy, Basic Energy Sciences, Materials Sciences and Engineering Division (DE-SC0008145 and DE-SC0012509) and AFOSR (FA9550-14-1-0277). X.X. thanks support from Cottrell Scholar Award and support from the State of Washington funded Clean Energy Institute.}

\section*{Acknowledgments}

\textbf{The authors thank Tobias Korn, Bernhard Urbaszek, and Hui Zhao for providing experimental data for this article.}

\bibliography{Ref}

\end{document}